\documentclass[12pt]{l4dc2023}

\title[Learning Disturbances Online for Risk-Aware Control]{Learning Disturbances Online for Risk-Aware Control: \\
Risk-Aware Flight with Less Than One Minute of Data}
\usepackage{times}
\usepackage{gen_settings}

\author{%
 \Name{Prithvi Akella$^1$} \Email{pakella@caltech.edu}\\
 \Name{Skylar X. Wei$^1$} \Email{swei@caltech.edu}\\
 \Name{Joel W. Burdick$^1$} \Email{jwb@robotics.caltech.edu}\\
 \Name{Aaron D. Ames$^1$} \Email{ames@caltech.edu}\\
 \addr $^1$1200 E California Blvd MC 104-44, Pasadena, CA 91101 
}

\begin{document}

\maketitle

\begin{abstract}%
Recent advances in safety-critical risk-aware control are predicated on \textit{apriori} knowledge of the disturbances a system might face.  This paper proposes a method to efficiently learn these disturbances online, in a risk-aware context.
First, we introduce the concept of a \textit{Surface-at-Risk}, a risk measure for stochastic processes that extends Value-at-Risk --- a commonly utilized risk measure in the risk-aware controls community.  Second, we model the norm of the state discrepancy between the model and the true system evolution as a scalar-valued stochastic process and determine an upper bound to its \textit{Surface-at-Risk} via Gaussian Process Regression.  Third, we provide theoretical results on the accuracy of our fitted surface subject to mild assumptions that are verifiable with respect to the data sets collected during system operation.  Finally, we experimentally verify our procedure by augmenting a drone's controller and highlight performance increases achieved via our risk-aware approach after collecting less than a minute of operating data.
\end{abstract}


\section{Introduction}
The models we use for control synthesis are useful, though oftentimes inaccurate. To wit, reduced order models are heavily utilized for controller synthesis for complex robotic systems, \textit{e.g.} quadrupeds, bipeds, drones, \textit{etc}~(\cite{bouman2020autonomous,fan2021step,ubellacker2021verifying,xiong2021reduced}).  However, these models require robustification to disturbances (e.g. to compensate for the gap between the reduced and full order models) to function reliably on these complex systems~(\cite{thieffry2018control,kim2020safe,alan2021safe,kolathaya2018input,ahmadi2020risk}).  As a result, recent studies on the robust control of nonlinear systems center around input-to-state-safe control~(\cite{kolathaya2018input,romdlony2016new,taylor2020learning}) and risk-aware control~(\cite{ahmadi2020risk,lindemann2021reactive,majumdar2020should,dixit2021risk,akella2022sample}) among other techniques.  These methods typically assume \textit{apriori} knowledge of a model and possible disturbances (or at least the magnitude thereof) and employ control techniques designed to reject those known disturbances.  On the other hand, learning-based approaches attempt to identify the underlying model~(\cite{buisson2020actively,nguyen2011model,jain2018learning,berkenkamp2015safe,folkestad2022koopnet,westenbroek2021combining,wang2018safe}), in many cases through Gaussian Process Regression (GPR)~(\cite{williams2006gaussian}).

However, assuming \textit{apriori} knowledge of disturbances might not be accurate in real-world settings, and gaussian process regression for model determination tends to be sample-complex and only uncover expected system behavior.  While learning expected behavior is indeed useful, control predicated on expected models of system behavior might yield problematic behavior in safety-critical settings where risk-sensitive approaches are preferable~(\cite{ahmadi2021constrained,ono2018mars}).  Skipping the model identification step, recent work in Bayesian Optimization and Reinforcement Learning aims to identify such risk-aware policies in a model-free fashion~(\cite{cakmak2020bayesian,makarova2021risk,heger1994consideration,chow2017risk,mihatsch2002risk,geibel2005risk}).  However, these prior works assume an ability to sample disturbances directly, assume \textit{apriori} knowledge of disturbances, or are sample-complex.

\begin{figure}[t]
    \centering
    \includegraphics[width = \textwidth]{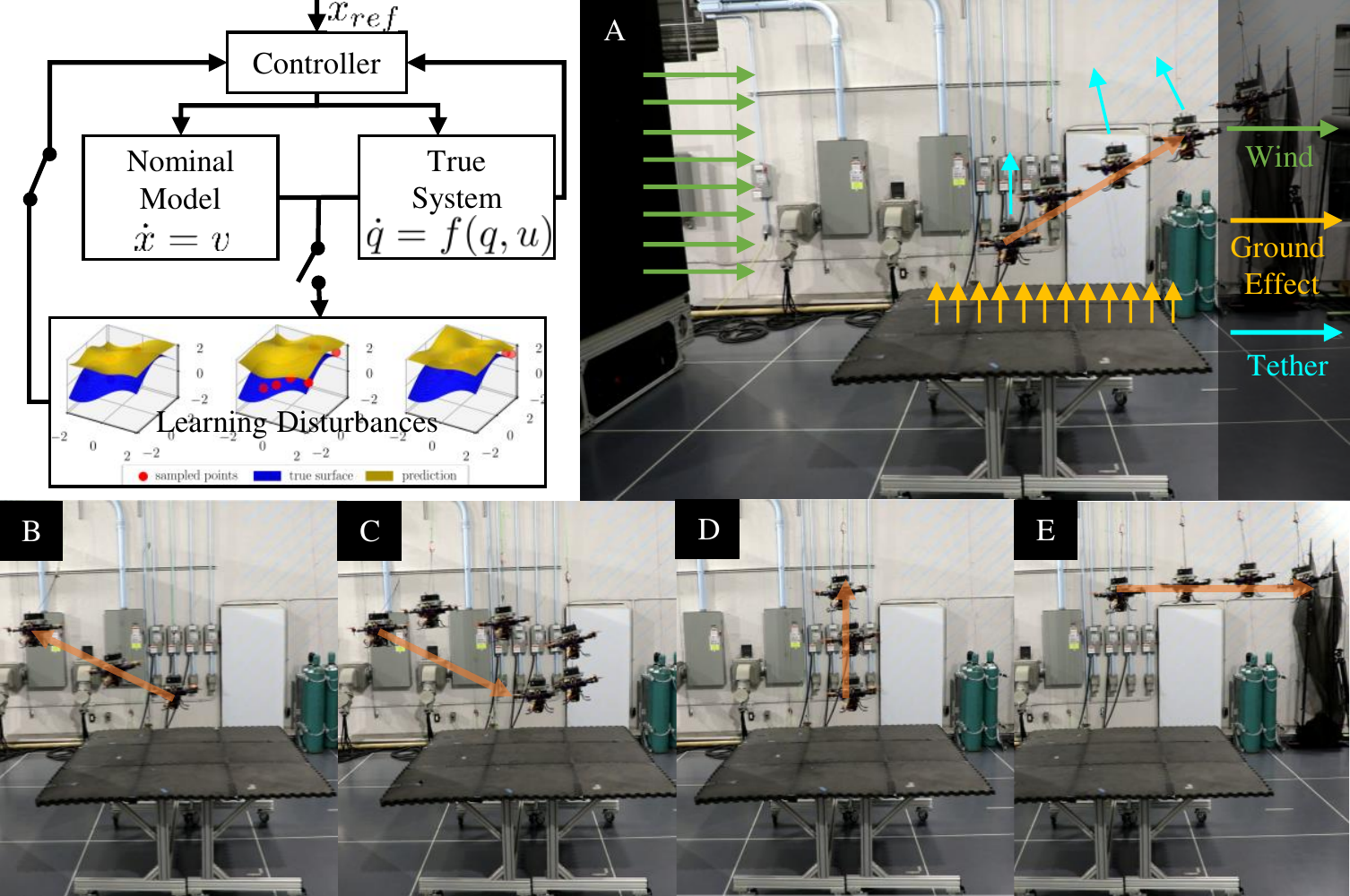}
    \caption{(Top Left) A general overview of our procedure, (Top Right) a photo of our experimental setup, and (Bottom) snippets of flight paths taken by the drone during the second set of experiments run --- the experiments depicted on the left in Figure~\ref{fig:experiment_breakdown}.  Our procedure has two parts.  First, we implement a nominal controller and calculate norm discrepancies between predicted model evolution and true system evolution.  Then, we fit, via gaussian process regression, a risk-aware disturbance model for the disturbances that the nominal system experiences.  We show in Section~\ref{sec:experimentation} how our procedure dramatically improves baseline controller performance and provide a statement on the theoretical accuracy of our model in Section~\ref{sec:learning_models}.}
    \vspace{-0.2 in}
    \label{fig:title_fig}
\end{figure}

\newidea{Our Contribution:} We propose a risk-aware model augmentation approach via learning disturbance models online that does not require \textit{apriori} disturbance knowledge.  Our approach is sample-efficient as shown in Section~\ref{sec:experimentation}, where we require less than a minute of flight data to make risk-aware control improvements on a drone mid-flight.  Furthermore, by building off prior work~(\cite{akella2022safety,akella2022sample}), we both define and ensure that our learned disturbance surface is a \textit{Surface-at-Risk} for the stochastic process accounting for the discrepancy between model and true system evolution.  Hence, augmenting the controller with our learned disturbance model yields an efficient risk-aware controller as we demonstrate experimentally.

\newidea{Structure:} Section~\ref{sec:background_gpr} provides a brief background on gaussian process regression, and Section~\ref{sec:VaR_surface} formally defines a \textit{Surface-at-Risk} for a stochastic process.  Section~\ref{sec:learning_models} presents the problem of upper-bounding such a surface and provides a theoretical statement on the accuracy of our procedure with respect to identifying such an upper bound.  Finally, Section~\ref{sec:experimentation} showcases the utility of our procedure for risk-aware control of a drone with online disturbance learning.

\section{Mathematical Preliminaries and Definitions}
\subsection{A Brief Aside on Gaussian Process Regression}
\label{sec:background_gpr}
A key concept in our approach is the notion of \textit{Surfaces-at-Risk} which we fit via GPR as part of our procedure.  GPR typically assumes the existence of an unknown function $f:X \to \mathbb{R}$ that we aim to represent by taking noisy samples $y$ of $f$ at points $x \in X$ where the noise $\xi$ is typically assumed to be sub-Gaussian~(\cite{srinivas2009gaussian,chowdhury2017kernelized, williams2006gaussian}).  Let $\GPX = \{x_i\}_{i=1}^N$ be a set of $N$ points $x \in X$ and $\GPY$ be the corresponding set of noisy observations, \textit{i.e.} $\GPY = \{y_i = f(x_i) + \xi,~\forall~x_i \in \GPX\}$.  Furthermore, let $k: X \times X \to \mathbb{R}$ be a positive-definite \textit{kernel function}.  Then, a \textit{gaussian process} is uniquely defined by its mean function $\mu: X \to \mathbb{R}$ and its variance function $\sigma: X \to \mathbb{R}$.  These functions are defined as follows, with $k_N(x) = [k(x,x_i)]_{x_i \in \GPX}$, $\mathbb{K} = [k(x_i,x_j)]_{x_i,x_j \in \GPX}$, $y_{1:N} = [y_i]_{y_i \in \GPY}$, and $\lambda = (1+\frac{2}{N})$:
\begin{align}
    \mu_N(x) = k_N(x)^T\left(\mathbb{K} + \lambda I_{N}\right)^{-1}y_{1:N}, \quad \sigma_N(x) = k_N(x,x), \label{eq:gp_functions}\\
    k_N(x,x') = k(x,x') - k_N(x)^T\left(\mathbb{K}_N + \lambda I\right)^{-1}k_N(x').
\end{align}

Lastly, each kernel function has a space of functions it can reproduce to point-wise accuracy, it's Reproducing Kernel Hilbert Space (RKHS).  Under the assumption that the function to-be-fitted $f$ has bounded norm in the RKHS of the chosen kernel $k$, GPR guarantees high-probability representation of $f$ as formalized in the theorem below, taken from~\cite{chowdhury2017kernelized}:
\setcounter{theorem}{0}
\begin{theorem}
    \label{thm:gp_concentration_inequalities}
    Let $f:X \to \mathbb{R}$, $\GPX = \{x_i\}_{i=1}^N$ be a set of $N$ points $x \in X$, $\GPY = \{y_i = f(x_i) + \xi\}_{x_i \in \GPX}$ be a set of noisy observations $y_i$ of $f(x_i)$ with $R$ sub-gaussian noise $\xi$, and $k: X \times X \to \mathbb{R}$ be a positive-definite kernel function.  If $f$ has $B$-bounded RKHS norm for some $B > 0$, \textit{i.e.} $\|f\|_{RKHS} \leq B$, then, with $\mu_N$ and $\sigma_N$ as per~\eqref{eq:gp_functions} and with minimum probability $1-\delta$,
    \begin{equation}
        |\mu_N(x) - f(x)| \leq \left(B + R \sqrt{2 \ln{\frac{\sqrt{\det\left((1+\frac{2}{N})I_N + \mathbb{K}_N\right)}}{\delta}}}\right) \sigma_N(x),~\forall~x \in X.
    \end{equation}
\end{theorem}

\subsection{Surfaces-at-Risk for Scalar Stochastic Processes}
\label{sec:VaR_surface}
This section formally defines a \textit{Surface-at-Risk} for a scalar stochastic process --- the specific structure we aim to fit via GPR.  Given a probability space $(\Omega, \mathcal{F}, \mathbb{P})$ with $\Omega$ a sample space, $\mathcal{F}$ a $\sigma$-algebra over $\Omega$ defining events, and $\mathbb{P}$ a probability measure, we define a scalar \textit{stochastic process} $S$ over the indexed space $\mathcal{X}$ as a collection of scalar random variables $S_x:\Omega \to \mathbb{R}$, \textit{i.e.} $S = \{S_x\}_{x \in \mathcal{X}}$.  Here, each scalar random variable $S_x$ has a (perhaps) different distribution $\pi_x:\mathbb{R} \to [0,1]$ such that probability of $S_x$ taking values in $A\subseteq \mathbb{R}$, \textit{i.e.} $\prob_{\pi_x}[S_x \in A \subseteq \mathbb{R}]$, is well-defined.

\begin{figure}[t]
    \centering
    \includegraphics[width = \textwidth]{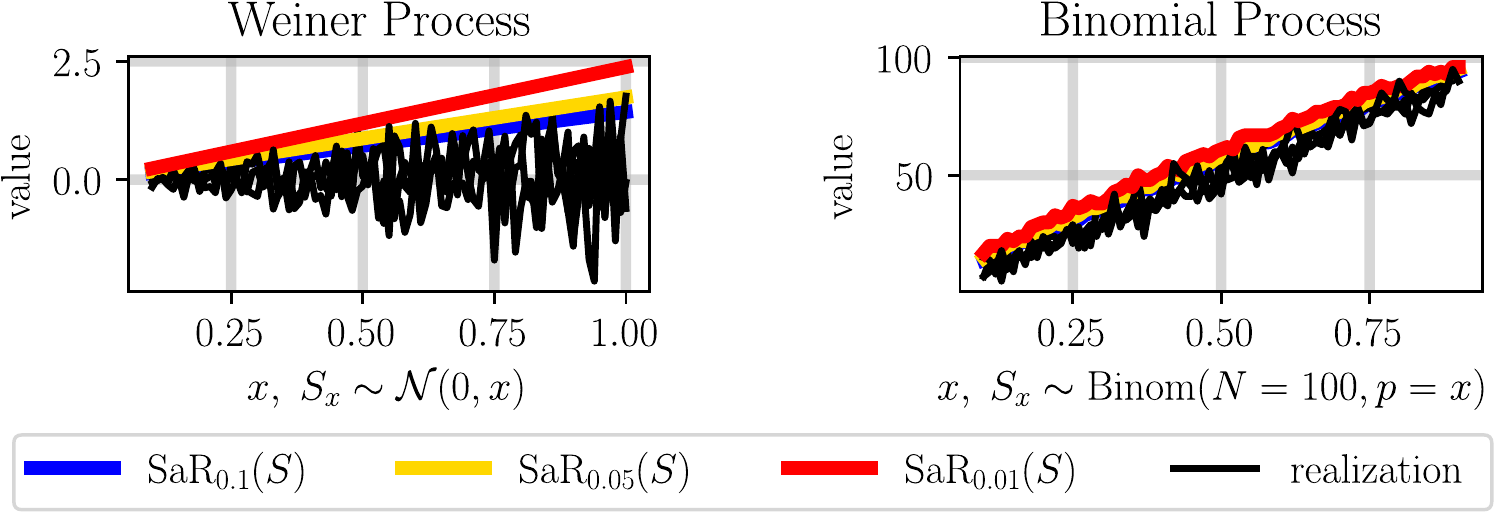}
    \caption{Example Surfaces-at-Risk at risk-levels $\epsilon \in [0.1,0.05,0.01]$ for a Weiner Process (Left) and Binomial Process (Right).  Distributions for the indexed scalar random variables $S_x$ comprising each process $S$ are provided on the axes. Sample realizations of the stochastic processes are shown in black, with Surfaces-at-Risk shown via colored lines.}
    \vspace{-0.2 in}
    \label{fig:sar_examples}
\end{figure}

Risk-measures are functions of these scalar random variables, and Value-at-Risk is a specific type of risk-measure stemming from the financial literature~(\cite{linsmeier2000value}).
\begin{definition}
    \label{def:value-at-risk}
    The \underline{Value-at-Risk} level $\epsilon \in [0,1]$ of a scalar random variable $X$ defined over the probability space $(\Omega,\mathcal{F},\mathbb{P})$ with distribution $\pi$ is defined as the $(1-\epsilon)-th$ quartile of $X$, \textit{i.e.}
    \begin{equation}
        \var_{\epsilon}(X) \triangleq c\, \suchthat\, c = \inf\{z \in \mathbb{R}~|~ \prob_{\pi}[X\leq z] \geq 1-\epsilon\}.
    \end{equation}
\end{definition} 
\noindent Then, the \textit{Surface-at-Risk} for a scalar stochastic process is a similar collection of the Values-at-Risk of the underlying scalar random variables constituting the scalar stochastic process. 
\begin{definition}
    \label{def:surface-at-risk}
    The \underline{Surface-at-Risk} level $\epsilon \in [0,1]$ of a scalar stochastic process $S$ indexed by the set $\mathcal{X}$ is the indexed collection of the Values-at-Risk level $\epsilon$ of each random variables $S_x$ comprising $S$:
    \begin{equation}
        \sar_{\epsilon}(S,x) = \var_{\epsilon}(S_x).
    \end{equation}
\end{definition}
\noindent Figure~\ref{fig:sar_examples} shows a few examples of Surfaces-at-Risk for varying risk-levels $\epsilon$ overlaid on realizations of common stochastic processes.

\section{Learning Disturbances}
\label{sec:learning_models}
\subsection{The Risk-Aware Disturbance-Norm Identification Problem}
\label{sec:problem_definition}
From a risk-aware standpoint, we aim to identify a \textit{Surface-at-Risk} as per Definition~\ref{def:surface-at-risk} for a scalar stochastic process $S$ indexed over the model state-space $\hat{\mathcal{X}}$.  Sample realizations of this process correspond to disturbance norms the system might experience at any given model state $\hat x \in \hat{\mathcal{X}}$. To formally state this problem, we will first denote our true system via $x$ and sim model via $\hat x$, \textit{i.e.} $\forall~k,j = 0,1,2,\dots$, (perhaps) different state and input spaces, and process noise $\xi$ with (unknown and perhaps) state-dependent distribution $\pi$
\begin{equation}
    \tag{SYS}
    \label{eq:systems}
    \begin{aligned}
    & \mathrm{\textbf{True:}}~~ & x_{k+1} & = f(x_k,u_k) + \xi, & x_k \in \mathcal{X},~u_k \in \mathcal{U},~\xi \sim \pi, \\
    & \mathrm{\textbf{Sim:}}~~ & \hat x_{j+1} & = \hat f(\hat x_j,\hat u_j), & \hat x_j \in \hat{\mathcal{X}},~\hat u_{j} \in \hat{\mathcal{U}}.
    \end{aligned}
\end{equation}
As an example consistent with the demonstration to follow, the true system would be a drone, with our reduced-order simulator model a single integrator.  The true state would be the drone's position and orientation, and the true input would be the rotor torques.  Meanwhile, the model state would be the drone's position in $3$-space, and the model input would be the desired velocity. 

To identify the discrepancy between the systems in~\eqref{eq:systems}, we define two maps - $M_x$ which projects the true state $x$ to the model state $\hat x$ and $M_u$ which extends the model input $\hat u$ to the true input $u$, \textit{e.g.} $M_u$ provides rotor torques to realize the desired velocity in $3$-space:
\begin{equation}
    \tag{MAPS}
    \label{eq:maps}
        M_x: \mathcal{X} \to \hat{\mathcal{X}}, \quad M_u: \hat{\mathcal{U}} \times \mathcal{X} \to \mathcal{U}.
\end{equation}
To note, we only assume the existence of these maps and the ability to use them, we do not assume that they are unique, we know their analytic form, \textit{etc}. To put these maps in the context of our drone example, the drone's underlying controller operates at $1$ kHz making the true-system time step $1$ ms.  Since we aim to provide model inputs at $50$ Hz, $K=20$.  $M_x$ is just the projection of our drone's position in $3$-space, and $M_u$ is the on-board controller that takes in a commanded $3$-space velocity --- model input $\hat u$ --- and updates rotor speeds at $1$ kHz to achieve that velocity.  These maps will be further explained in Section~\ref{sec:experimentation}.  Finally, we assume that after some amount of true system time-steps $K > 0$, we can observe projected true system evolution.  We denote $K$ as the \textit{time-dilation parameter} and the observation function $O$ is defined as follows:
\begin{equation}
    \tag{OBS}
    \label{eq:observation_function}
   x_{k+1} = f(x_k,M_u(\hat{u},x_k)),~ O(x_0,\hat{u}) = M_x(x_K).
\end{equation}

These maps let us formally state the projected evolution of our true system, \textit{i.e.} evolution of $\hat x_j = M_x(x_{Kj})$, when driven by a feedback controller $U: \hat{\mathcal{X}} \to \hat{\mathcal{U}}$.  Comparing projected and sim model evolution results in the discrepancy $d$ we aim to learn:
\begin{equation}
    \label{eq:disturbance_definition}
    \hat x_{j+1} = \hat f(M_x(x_{Kj}),U(M_x(x_{Kj})) + \underbrace{O(x_{Kj},U(M_x(x_{Kj})) - \hat f(M_x(x_{Kj}),U(M_x(x_{Kj}))}_{d,~\mathrm{and~}\dnormsample = \|d\| \mathrm{~has~distribution~} \pi_{\hat{x}}:\mathbb{R} \to [0,1]}.
\end{equation}
Then, inspired by input-to-state-safe barrier and input-to-state-stable Lyapunov works whose robust controllers only require information on the $2$-norm of this disturbance $d$, we aim to learn a probabilistic upper bound on $\|d\|$ by taking samples of indexed random variables $S_{\hat{x}}$ comprising a disturbance-norm stochastic process $S$ indexed by $\hat{\mathcal{X}}$ as in~\eqref{eq:systems}.
\begin{definition}
    \label{def:disturbance_stochastic_process}
    The \underline{disturbance-norm stochastic process} $S = \{S_{\hat x}\}_{\hat x \in \hat{\mathcal{X}}}$ where samples of each random variable $S_{\hat{x}}$ correspond to norms $\dnormsample$ of disturbances $d$ as defined in equation~\eqref{eq:disturbance_definition}.  The variability in norm samples $\delta$  arises through the assumed process noise $\xi$ in the true system dynamics in~\eqref{eq:systems}.
\end{definition}

\newidea{Remark on Residuals:} If we only consider a deterministic discrepancy between the true and sim models, then the disturbances $d$ as per~\eqref{eq:disturbance_definition} would correspond to residual dynamics, and our procedure would fit a surface to the norm of the residual dynamics (learning residual dynamics has a well-studied history, see \cite{saveriano2017data,johannink2019residual,schperberg2022real,zeng2020tossingbot} and citations within).  The discrepancy between these approaches and ours is that we also learn a probabilistic bound on the norm of any stochastic, model-state-dependent disturbances that affect the system during operation.  This is why we represent the discrepancies as a stochastic process and fit a Surface-at-Risk, which provides a natural way to reason about risk-aware disturbance rejection in a context including model errors and stochastic uncertainty.

Furthermore, we assume our disturbance-norm stochastic process is indexed over the model state space $\hat{\mathcal{X}}$ as opposed to the true state space $\mathcal{X}$ as we only assume the ability to measure the projected state $\hat x_j = M_x(x_{Kj})$.  Therefore, we can only correspond sampled disturbance norms $\dnormsample$ to points in the projected state space $\hat{\mathcal{X}}$.  Then, our goal is to identify a ``close" upper bound to the \textit{Surface-at-Risk} for this disturbance-norm stochastic process at some risk-level $\epsilon \in [0,1]$.
\begin{problem}
    \label{problem_statement}
    Identify an upper bound to the \textit{Surface-at-Risk} at some risk-level $\epsilon \in [0,1]$ for the disturbance-norm stochastic process $S$ as per Definition~\ref{def:disturbance_stochastic_process} with Surfaces-at-Risk as defined in Definition~\ref{def:surface-at-risk}.  Specifically, identify an estimate $\sarestimate_{\epsilon}$ such that,
    \begin{equation}
        \label{eq:upper_bounding_surface}
        \sarestimate_{\epsilon}(S,\hat{x}) \geq \sar_{\epsilon}(S,\hat{x}),~\forall~\hat x \in \hat{\mathcal{X}}.
    \end{equation}
\end{problem}
While the aforementioned upper bound $\sarestimate_{\epsilon}$ could be arbitrarily large and satisfy~\eqref{eq:upper_bounding_surface}, we aim to find a ``close" upper bound to the true Surface-at-Risk level $\epsilon$ to facilitate risk-aware control.

\subsection{Fitting a Disturbance-Norm Surface-at-Risk}
\label{sec:fitting_surfaces}
For identifying such an upper bound~$\sarestimate_{\epsilon}$, we first note that even for stochastic processes whose sample realizations are non-differentiable, their Surfaces-at-Risk are relatively smooth --- see Figure~\ref{fig:sar_examples} for examples.  Intuitively, we expect the disturbance norms $\dnormsample_i,\dnormsample_j$ at ``close" model states $\hat{x}_i,\hat{x}_j \in \hat{\mathcal{X}}$ are similarly ``close":
\begin{assumption}
    \label{assump:bounded_discrepancy}
    For the disturbance-norm stochastic process $S$ in Definition~\ref{def:disturbance_stochastic_process}, the \textit{Surface-at-Risk} at a given risk-level $\epsilon \in [0,1]$ has bounded discrepancy. \textit{I.e.} $\exists~\alpha,\beta \in \mathbb{R}_{\geq 0}$ such that,
    \begin{equation}
        \forall~\hat{x}_i, \hat{x}_j \in \hat{\mathcal{X}},~\|\hat{x}_i - \hat{x}_j\| \leq \alpha \implies |\sar_{\epsilon}(S,\hat{x}_i) - \sar_{\epsilon}(S,\hat{x}_j)| \leq \beta.
    \end{equation}
\end{assumption}
Notably, this assumption only implies a bounded discrepancy, and not continuity, \textit{e.g.} a bounded piecewise continuous function would have bounded variance as per our assumption. We will verify that this assumption holds for the data set we collect in Section~\ref{sec:experimentation}.

Second, we need to take (perhaps noisy) unbiased samples of $\sarestimate_{\epsilon}(S,\hat{x})$ for a given model state $\hat{x} \in \hat{\mathcal{X}}$.  By equation~\eqref{eq:upper_bounding_surface}, $\sarestimate_{\epsilon}(S,\hat{x}) \geq \var_{\epsilon}(S_{\hat {x}})$, and we can define one sample $\dnormsample_j$ of $S_{\hat{x}_j}$ as follows, where $O$ is as per~\eqref{eq:observation_function}, and $M_x$ is as per~\eqref{eq:maps}:
\begin{equation}
    \label{eq:norm_sample_definition}
    \dnormsample_j = \|O(x_{Kj},U(M_x(x_{Kj})) - \hat f(M_x(x_{Kj}),U(M_x(x_{Kj}))\|,\quad \hat{x}_j = M_x(x_{Kj}).
\end{equation}
Then, we can group multiple samples $\dnormsample_j$ for sequential model states visited during operation, \textit{i.e.} $\dnormsample_j, \dnormsample_{j+1}, \dots$ for $\hat{x}_j,\hat{x}_{j+1},\dots$ to produce an upper bound to at least one Value-at-Risk level $\epsilon$ of a sampled random variable, \textit{i.e.} $\var_{\epsilon}(S_{\hat{x}_j}), \var_{\epsilon}(S_{\hat{x}_{j+1}}), \dots$.  To do so, we require the following theorem, stated for $N$ scalar random variables $X$ with (perhaps) different distributions $\pi$.
\setcounter{theorem}{1}
\begin{proposition}
    \label{prop:group_samples}
    Let $\{X_i\}_{i=1}^N$ be a collection of $N$ scalar random variables with (perhaps) different distributions $\{\pi_i\}_{i=1}^N$, and let $\{x_i\}_{i=1}^N$ be a set of $N$ samples of these random variables, one sample per each random variable, \textit{i.e.} $x_i$ is a sample of $X_i$.  Then, for any $\epsilon \in [0,1]$, the probability that at least one sample $x_{\ell} \in \{x_i\}_{i=1}^N$ is greater than the Value-at-Risk level $\epsilon$ of its corresponding random variable $X_{\ell}$ is equivalent to $1-(1-\epsilon)^N$, \textit{i.e.} with $\var$ as per Definition~\ref{def:value-at-risk} and $\forall~\epsilon \in [0,1]$,
    \begin{equation}
        \prob_{\pi_1,\pi_2,\dots,\pi_N}\left[\exists~x_{\ell} \in \{x_i\}_{i=1}^N \suchthat x_{\ell} \geq \var_{\epsilon}(X_{\ell}) \right] \geq 1-(1-\epsilon)^N.
    \end{equation}
\end{proposition}
\begin{proof}
    Consider a random variable $X_{\ell} \in \{X_i\}_{i=1}^N$.  The probability of taking a sample $x_{\ell}$ of $X_{\ell}$ such that $x_{\ell} \geq \var_{\epsilon}(X_{\ell})$ is less than or equal to $\epsilon$ by Definition~\ref{def:value-at-risk}.  The same line of reasoning holds $\forall~X_{\ell} \in \{X_i\}_{i=1}^N$.  As such, the probability that no sample $x_{\ell} \in \{x_i\}_{i=1}^N$ is greater than the corresponding Value-at-Risk level $\epsilon$ is less than or equal to $(1-\epsilon)^N$, yielding our result.
\end{proof}

\begin{algorithm}[t]
\SetAlgoLined

\hrulefill
\caption{Fitting a Disturbance-Norm Surface-at-Risk} \vspace{-0.1 in} \label{alg:fitting_surface}
\hrulefill

\KwData{$\alpha,\beta$ for Assumption~\ref{assump:bounded_discrepancy}, an integer $\sampleamount>0$ for Proposition~\ref{prop:group_samples} corresponding to the number of random variables to sample, time-step dilation parameter $K > 0$ between true system evolution and model evolution as per~\eqref{eq:observation_function}, and $k:\hat{\mathcal{X}} \times \hat{\mathcal{X}} \to \mathbb{R}$ a kernel function}
\textbf{Initialize:}  $\iter = 0$, $\GPX = $ [], $\GPY = $ [] \;

\textbf{References:} Disturbance Norm samples $\dnormsample_j$ as per~\eqref{eq:norm_sample_definition} and projector $M_x$ as per~\eqref{eq:maps} \;

\While{True}{
Initialize empty data-set, \textit{i.e.} $\mathcal{D}_{\iter} = $ [ ] \;

\For{$j=\sampleamount \cdot \iter, \sampleamount \cdot \iter +1, \dots, \sampleamount (\iter +1) - 1$}{
Collect state-indexed disturbance norm samples, \textit{i.e.} $\mathcal{D}_{\iter} \leftarrow \mathcal{D}_{\iter} \cup (\dnormsample_j,\hat{x}_j = M_x(x_{Kj}))$\;
}
Augment GP state dataset with $\mathcal{D}_{\iter}$:~$\GPX \leftarrow \GPX \cup~\hat{x}_{\sampleamount(\iter+1)-1}$\;

Augment GP norm dataset with $\mathcal{D}_{\iter}$:~ $\GPY \leftarrow \GPY \cup~\max\{\dnormsample_{\ell} \in \mathcal{D}\} + \beta$ \;

Fit $\mu_{\iter},\sigma_{\iter}$ as per~\eqref{eq:gp_functions} with data sets $\GPX,\GPY$.  $\iter ++$ \;
}
\nonl \hrulefill
\end{algorithm}

Our procedure for generating unbiased samples of the upper bound~$\sarestimate_{\epsilon}$ stems directly from Proposition~\ref{prop:group_samples} and Assumption~\ref{assump:bounded_discrepancy}.  First, we let the system evolves for $\sampleamount$ model time-steps and collect one norm sample $\dnormsample_j$ per model state $\hat{x}_j$ visited during operation.  This norm sample $\dnormsample_j$ is calculated as per~\eqref{eq:norm_sample_definition}.  Second, Proposition~\ref{prop:group_samples} guarantees that the largest norm sample $\dnormsample^*_j$ is greater than the Value-at-Risk level $\epsilon$ for its corresponding indexed random variable $S_{\hat{x}^*_j}$ with some minimum probability.  Third, if all norm samples were drawn from indexed random variables $S_{\hat{x}_j}$ whose indices $\hat{x}_j$ were ``close", \textit{i.e.} $\|\hat{x}_s - \hat{x}_r\| \leq \alpha$ $\forall~\hat x_r \neq \hat x_s \in \{\hat{x}_{j+i}\}_{i=0}^{N-1}$ and for some $\alpha > 0$, we can use Assumption~\ref{assump:bounded_discrepancy} to augment the largest norm sample $\dnormsample^*_j$ by a constant $\beta > 0$.  The sum is, with minimum probability $1-(1-\epsilon)^N$, an unbiased, non-noisy sample of $\sarestimate_{\epsilon}(S,\hat{x}_j)$.  Algorithm~\ref{alg:fitting_surface} formalizes this procedure and our main theoretical result follows.
\setcounter{theorem}{1}
\begin{theorem}
    \label{thm:main_theorem}
    Let $\alpha,\beta,\sampleamount,\iter,\mu_{\iter},\sigma_{\iter},$ and $k$ be as defined in Algorithm~\ref{alg:fitting_surface}, let $B > 0$, let $\sar$ be the \textit{Surface-at-Risk} measure as per Definition~\ref{def:surface-at-risk} for some risk-level $\epsilon \in [0,1]$, let $S$ be the disturbance-norm stochastic process as per Definition~\ref{def:disturbance_stochastic_process}, and let Assumption~\ref{assump:bounded_discrepancy} hold for each data set $\mathcal{D}_{\iter}$ in lines~5-7 of Algorithm~\ref{alg:fitting_surface} with respect to the given parameters $\alpha,\beta$.  If $\|\sarestimate_{\epsilon}(S)\|_{RKHS} \leq B$, then with minimum probability $\left(1-(1-\epsilon)^{\sampleamount}\right)^{\iter}$ the following holds $\forall~\hat{x} \in \hat{\mathcal{X}}$ and $\forall~\iter = 1,2,\dots$:
    \begin{equation}
        |\mu_{\iter}(\hat{x}) - \sarestimate_{\epsilon}(S,\hat{x})| \leq B \sigma_{\iter}(\hat{x}), \qquad \mu_{\iter}(\hat{x}) + B \sigma_{\iter}(\hat{x}) \geq \sar_{\epsilon}(S,\hat{x}).
    \end{equation}
\end{theorem}
\begin{proof}
First, by the assumptions above, we know that for each data set $\mathcal{D}_{\iter}$ in lines~5-7 of Algorithm~\ref{alg:fitting_surface}, we have taken one sample $\dnormsample_j$ of $\sampleamount$ (potentially) different random variables $S_{\hat{x}_j}$.  By Proposition~\ref{prop:group_samples}, we know that with minimum probability $1-(1-\epsilon)^{\sampleamount}$, the maximum sample $\dnormsample^*_j \triangleq \max\{\dnormsample_{\ell} \in \mathcal{D}\}$ is greater than the Value-at-Risk of its corresponding random variable $\var_{\epsilon}(S_{\hat{x}^*_j})$ ($\var$ is defined in Definition~\ref{def:value-at-risk}).  Since we assume Assumption~\ref{assump:bounded_discrepancy} holds for each such set of random variables, then we know that with minimum probability $1-(1-\epsilon)^{\sampleamount}$, the sum $\dnormsample^*_j + \beta$ is greater than the value-at-risk level $\epsilon$ of any sampled random variable, \textit{i.e.} the sum $\dnormsample^*_j + \beta$  is a non-noisy estimate of $\sarestimate_{\epsilon}(S,\hat{x}),~\forall~\hat{x} \in \mathcal{D}_{\iter}$.  Hence, repeating this same argument for each data point in $\GPX,\GPY$  and setting $R=0$, as each sampled point is a non-noisy sample of our upper-bounding surface, we recover the results of Theorem~\ref{thm:gp_concentration_inequalities} with minimum probability $(1-(1-\epsilon)^{\sampleamount})^{\iter}$:
\begin{equation}
    \label{eq:gp_sar_estimate}
    |\mu_{\iter}(\hat{x}) - \sarestimate_{\epsilon}(S,\hat{x})| \leq B \sigma_{\iter}(\hat{x}),~\forall~\hat{x} \in \hat{\mathcal{X}}.
\end{equation}

\noindent Our final result holds by unraveling the absolute-value inequality in~\eqref{eq:gp_sar_estimate}, as $\sarestimate_{\epsilon}(S)$ is an upper-bounding surface for $\sar_{\epsilon}(S)$.
\end{proof}

\section{Learning Disturbances Mid-Flight for Risk-Aware Control}
\label{sec:experimentation}

\begin{figure}[t]
    \centering
    \includegraphics[width = \textwidth ]{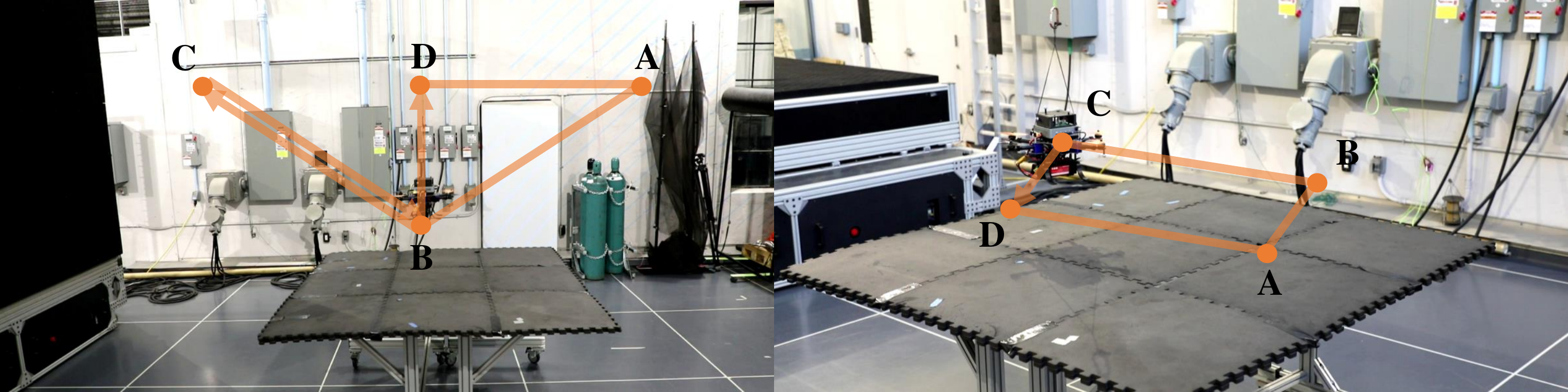}
    \caption{Depictions of the two types of periodic trajectories implemented in our drone experiments described in Section~\ref{sec:experimentation}.  These trajectories approximate difficult types of behaviors commonly asked of drones, 
    }
    \vspace{-0.2 in}
    \label{fig:experiment_breakdown}
\end{figure}


\subsection{Implementation Specifics}
\label{sec:implementation}
All flight tests are performed at the Caltech Center for Autonomous Systems and Technology arena which is equipped with an Optitrack motion capture system that samples and streams the rotor-craft pose at 190 Hz. We belay a safeguard tether to the drone (weights 2.46 kg) with a $\sim\!\!200\,$g passive weight attached on the other end to partially eliminate tether slack, which is another source of uncertainty. Figure~\ref{fig:experiment_breakdown} depicts the two types of flight paths taken, wherein we aimed to realize complex behaviors commonly asked of drones, \textit{e.g.} ascent and descent with both headwind and tailwind, circulating low to the ground, and taking off vertically in the presence of transverse wind.  All disturbing winds were realized by The Caltech Real Weather Wind Tunnel, and windspeed information was not made available to the baseline controller to-be-augmented.  This baseline controller was developed against a single integrator model, and as such, it outputs $3$-space velocities at $50$ Hz for the drone to follow.  The velocities provided by this controller are tracked by the drone's onboard flight controller, a Hex Cube Orange running a PX4 autopilot \cite{px4}.

With respect to the mathematical setting in Section~\ref{sec:problem_definition} then, we do not know our true system dynamics, though we model the system as a single integrator:
\begin{equation}
    \tag{EXP-SYS}
    \label{eq:experiment_systems}
    \hat x_{j+1} = \hat{x}_j + \hat{u}_j(\Delta t = 0.02),~\hat x_j \in \underbrace{[-2,2]^2 \times [1.2,2]}_{\hat{\mathcal{X}}},~\hat u_{j} \in \underbrace{[-0.8,0.8]^2 \times [-0.5,0.5]}_{\hat{\mathcal{U}}}.
\end{equation}
The state projection map $M_x$ as in~\eqref{eq:maps} reads the drone's position in $3$-space.  The input map $M_u$ corresponds to the onboard PX4 controller that maps true drone states $x \in \mathcal{X}$ and commanded $3$-space velocities $\hat{u} \in \hat{\mathcal{U}}$ to rotor speeds at $1$ kHz.  As we update these desired velocities at $50$ Hz, our time-dilation parameter $K = 20$ for Algorithm~\ref{alg:fitting_surface}.  Finally, our observation function $O$ as per~\eqref{eq:observation_function} outputs the projected true-system $3$-space position after $K$ true-system time-steps, and our disturbance-norm samples $\dnormsample$ as per~\eqref{eq:norm_sample_definition} are defined as follows:
\begin{equation}
    \label{eq:experiment_norm_sample}
    \dnormsample_j = \|O(x_{Kj},U(M_x(x_{Kj})) - (\hat{x}_j + U(M_x(x_{Kj}))\Delta t\|,\quad \hat{x}_j = M_x(x_{Kj}).
\end{equation}
The baseline controller $U: \hat{\mathcal{X}} \to \hat{\mathcal{U}}$ is a discrete-time Lyapunov controller designed to send the single-integrator system to a provided waypoint, and does not take into account complex aerodynamic effects, \textit{e.g.} ground effects, transverse wind, and tethered disturbances, which are challenging to model and can degrade flight performance when ignored (\cite{neuralfly,folkestad2022koopnet}). Furthermore, the number of random variables sampled per data-collection step $\sampleamount = 60$ was kept constant, and we used the squared-exponential kernel function with length-scale parameter $\ell = 1.0$ for all experiments as well.

Our desired outcomes were twofold.  First, we fit an upper bound to the disturbance-norm \textit{Surface-at-Risk} level $\epsilon = 0.05$ over the course of one traversal of the desired flight path.  In this initial flight path, we only implement the baseline controller and augment this controller if the system takes longer than $10$ seconds to reach within $0.1$ m of the subsequent waypoint along the desired path. As each path comprises fewer than $6$ waypoints, this ensures that our learned model considers less than a minute of data for all experiments on both flight paths.  These cutoff times were specifically chosen to highlight the efficiency of our method with limited data.  Second, on all subsequent flight paths, we provide from our fitted surface the norm of disturbances that the Lyapunov controller should reject while providing velocity commands.  As such, we expect performance improvements from our augmented controller in the form of traversal time speedups through the series of waypoints, as subsequent waypoints are provided once the drone reaches within $0.1$ m of the current, commanded waypoint, and the drone's controller should account for the vast majority of disturbances caused by wind, ground, and tether effects as we fitted an upper bound to the disturbance-norm \textit{Surface-at-Risk} level $\epsilon = 0.05$.

\subsection{Discussion of Results}
\label{sec:discussion}
We performed four sets of experiments: (A) Hovering and moving while maintaining a $0.15$ m height above ground (see right in Figure~\ref{fig:experiment_breakdown}); (B) Ascent, descent, and vertical take-off without any wind (see left in Figure~\ref{fig:experiment_breakdown}); (C) The same flight path as (B) but with a $0.6$ m/s transverse wind. The wind flows from left to right when looking at the setup in Figure~\ref{fig:experiment_breakdown}.  A graphical example is also provided in Figure~\ref{fig:title_fig}; (D) The same flight path as (B) and (C) but with a $2$ m/s transverse wind.

\begin{figure}[t]
    \centering
    \includegraphics{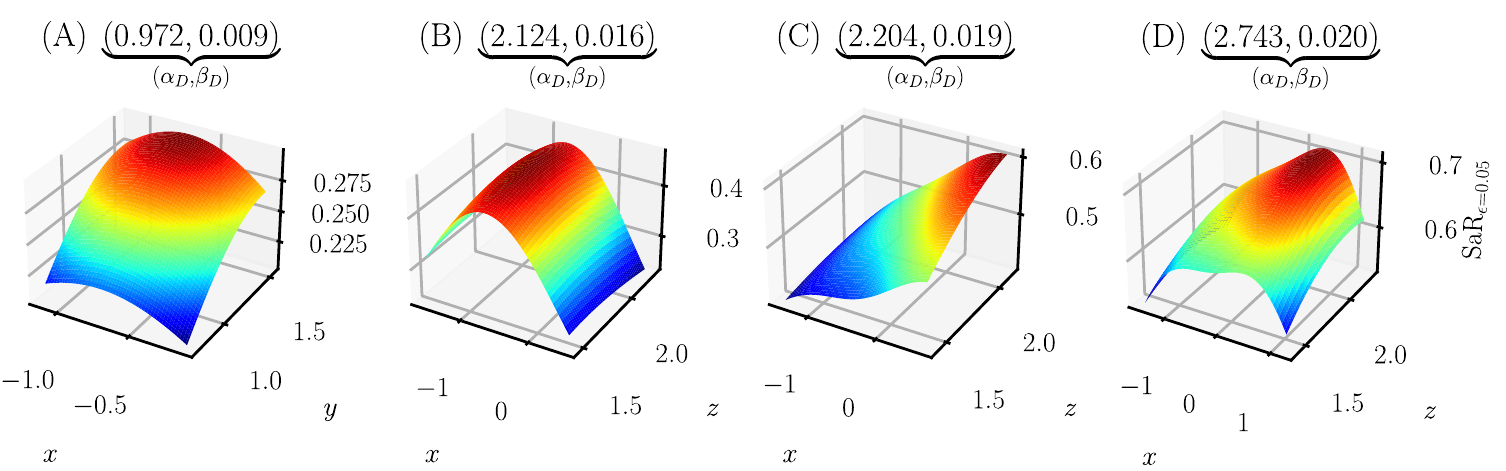}
    \caption{Fitted $\sar_{\epsilon=0.05}$ for the four experiments depicted in Figure~\ref{fig:experiment_breakdown}, with $\alpha_D$ the maximum distance between two sampled states for GPR, and $\beta_D$ the maximum discrepancy between two sampled disturbance norms.  Over all four experiments, we see a consistent $2 \times$ speedup in flight path times after implementation of the augmented controller --- a qualitative result we expect as per Theorem~\ref{thm:main_theorem}, as we fit an upper bound to disturbance norms at $95\%$ probability.  This information is further explained in Section~\ref{sec:discussion}.}
    \vspace{-0.2 in}
    \label{fig:data_compilation}
\end{figure}

Figure~\ref{fig:data_compilation} shows the fitted $\sar_{\epsilon = 0.05}$ for each of the four experiments (A)-(D) ran on the drone, as labeled prior.  As mentioned, in all cases we see at least a $2\times$ speedup in flight path times when implementing the augmented controller, with as much as a $5 \times$ speedup in the hovering case (A).  Furthermore, we were also able to verify Assumption~\ref{assump:bounded_discrepancy} with respect to the data sets we collected for each experiment.  Specifically, for (A), we assumed that for states within $\alpha = 1$m their Values-at-Risk level $\epsilon = 0.05$ would not change by more than $\beta = 0.05$.  As can be seen in the title of the associated subfigure in Figure~\ref{fig:data_compilation}, the reported values from data are smaller than their assumed counterparts, indicating that Assumption~\ref{assump:bounded_discrepancy} held over this experiment, at least with respect to the collected data.  For the remaining experiments, the assumed $\alpha,\beta$ values were as follows: (B) $\alpha = 3$m, $\beta = 0.05$; (C) $\alpha = 3$m, $\beta = 0.1$; (D) $\alpha = 3$m, $\beta = 0.2$.  Therefore, as we can see from the associated titles in Figure~\ref{fig:data_compilation}, we are similarly able to verify that Assumption~\ref{assump:bounded_discrepancy} held over each of these cases as well --- at least with respect to the data collected.  As such, we expected a significant increase in performance according to Theorem~\ref{thm:main_theorem} as was realized in all four cases with respect to flight path time speedups.  All experiments can also be seen in our supplementary video here: \cite{video}.

\section{Concluding Remarks and Future Work}
\label{sec:conclusion}
Our results were threefold.  First, we defined Surfaces-at-Risk, an extension of Value-at-Risk to scalar-valued stochastic processes.  Second, we defined the discrepancy between simulator and true-system evolution as a stochastic process, and provided a method to fit an upper bound to this process's \textit{Surface-at-Risk}.  Third, we provided a theoretical statement on the accuracy of our proposed approach with respect to fitting such an upper bound.  Finally, we showcased the utility of our procedure in facilitating risk-aware control by implementing our procedure on a drone mid-flight and exhibiting dramatic performance improvements as a result.  In future work, we hope to integrate our procedure with existing works in safety-critical control, to create a pipeline for online, adaptive, safety-critical risk-aware control.

\acks{The work of Prithvi Akella was supported by the Air Force Office of Scientific Research, grant FA9550-19-1-0302, and the National Science Foundation, grant 1932091.  The work of Skylar Wei was supported in part by DARPA, through the Learning and Introspective Control program. We would also like to thank the Caltech Center for Autonomous Systems and Technologies for the use of the wind tunnel in our experiments.  }

\bibliography{bib_works.bib}

\end{document}